*Review*

# AI in ESG for Financial Institutions: An Industrial Survey

**Jun Xu [1,*]**

[1]   ED of Machine Learning Engineering, A top internal bank

*   Correspondence: xujun@ieee.org

**Abstract:** The burgeoning integration of Artificial Intelligence (AI) into Environmental, Social, and Governance (ESG) initiatives within the financial sector represents a paradigm shift towards more sustainable and equitable financial practices. This paper surveys the industrial landscape to delineate the necessity and impact of AI in bolstering ESG frameworks. With the advent of stringent regulatory requirements and heightened stakeholder awareness, financial institutions (FIs) are increasingly compelled to adopt ESG criteria. AI emerges as a pivotal tool in navigating the complex interplay of financial activities and sustainability goals. Our survey categorizes AI applications across three main pillars of ESG, illustrating how AI enhances analytical capabilities, risk assessment, customer engagement, reporting accuracy and more. Further, we delve into the critical considerations surrounding the use of data and the development of models, underscoring the importance of data quality, privacy, and model robustness. The paper also addresses the imperative of responsible and sustainable AI, emphasizing the ethical dimensions of AI deployment in ESG-related banking processes. Conclusively, our findings suggest that while AI offers transformative potential for ESG in banking, it also poses significant challenges that necessitate careful consideration. The final part of the paper synthesizes the survey's insights, proposing a forward-looking stance on the adoption of AI in ESG practices. We conclude with recommendations with a reference architecture for future research and development, advocating for a balanced approach that leverages AI's strengths while mitigating its risks within the ESG domain.

## 1. Introduction

Environmental, Social, and Governance (ESG) criteria have become increasingly crucial in the financial industry, reflecting a growing awareness of the role financial institutions play in fostering sustainable development. This perspective is not only a response to evolving societal expectations but also a strategic approach to risk management and value creation in a rapidly changing world [1].

### 1.1 Three Pillars

#### 1.1.1 Environmental Considerations

FIs are increasingly recognizing the impact of environmental issues on their operations and the broader economy. Climate change, in particular, presents both direct and indirect risks. Direct risks include the physical impacts of climate events on bank assets and investments, while indirect risks involve the transition to a low-carbon economy, potentially stranding assets in carbon-intensive sectors. By integrating environmental criteria into their decision-making processes, banks can better manage these risks and capitalize on opportunities in green finance, such as renewable energy projects and sustainable infrastructure financing.

### 1.1.2 Social Factors

Social considerations encompass issues like labor practices, diversity and inclusion, and human rights. In the financial sector, social factors are gaining prominence due to their implications for reputational risk and customer loyalty. FIs are increasingly expected to ensure that their practices and those of their clients meet certain social standards. This shift is not just about risk mitigation; it also opens opportunities for banks to develop new products and services that address social challenges, such as financial inclusion and affordable housing.

### 1.1.3 Governance Aspects

Governance relates to how a company is managed, including its corporate governance practices, executive compensation, internal controls, and shareholder rights. For FIs, strong governance is crucial for maintaining trust and stability. It encompasses regulatory compliance, ethical conduct, and risk management. Effective governance can prevent malpractices and scandals, which can have severe financial and reputational consequences.

Table 1 describes the main considerations of each pillar with their typical examples and actions.

### 1.2 Necessity of ESG

The necessity of ESG in banking is multifaceted. First, there is growing regulatory pressure. Governments and international bodies are increasingly mandating ESG disclosures and integrating sustainability considerations into financial regulations. FIs that proactively adopt ESG principles are better positioned to comply with these evolving regulatory requirements.

Second, there is a clear economic rationale. ESG-focused banking can open new markets and revenue streams. Investments in sustainable projects and businesses are growing, driven by both consumer preferences and the recognition of the long-term economic benefits of sustainability. Researches show that the total assets of ESG funds showed significant growth, and various predictions indicate a continuing upward trend in the recent years. Bloomberg Intelligence stated that the global ESG assets were expected to surpass $41 trillion by 2022 and reach $50 trillion by 2025, accounting for one-third of the projected total global assets under management [2]. PwC Global said that the ESG-focused institutional investment assets under management (AuM) is expected to soar by 84% to $33.9 trillion in 2026, making up 21.5% of total assets under management [3]. Standard Chartered Bank discussed how $8.2 trillion of investable retail wealth could be channeled into sustainable investments by 2030 to finance ESG objectives in growth markets [4]. Harvard Law School Forum on Corporate Governance predicted that ESG AuM in the US is projected to more than double, from $4.5 trillion in 2021 to $10.5 trillion in the near future [5]. A Morgan Stanley survey states, "85% of US investors are interested in sustainable investing." [6]

Third, ESG is vital for risk management. Environmental risks, social unrest, or governance scandals can significantly impact a bank's financial performance. By incorporating ESG criteria, FIs can better anticipate and mitigate these risks.

Forth, we shall delve into the significance of corporate responsibility within a sustainable context. Historically, until the 1990s, a substantial portion of a company's value was attributed to "tangible" assets. Contrastingly, in contemporary times, the valuation of the world's leading corporations largely hinges on "intangible" assets. These encompass business concepts, goodwill, intellectual capital, patents, brands, and reputation. Unlike tangible assets, intangible assets are susceptible to impairment, potentially leading to a marked decrease in a company's value, especially if linked to adverse incidents. Such incidents often correlate with sustainability challenges, including exposure of substandard working conditions among suppliers in low-cost countries, environmental harm, and corruption. Conversely, when corporations adeptly portray themselves as sustainable and cultivate a brand that embodies sustainability principles, there is an enhancement in their intangible assets [7].

Lastly, there is a moral and ethical dimension. FIs, as central pillars of the economy, have a responsibility to support sustainable development and ethical practices. This responsibility is increasingly recognized by stakeholders, including customers, employees, and investors, who are demanding greater transparency and responsibility from financial institutions.

ESG considerations are not just a trend but a fundamental shift in the financial industry, driven by regulatory changes, economic opportunities, risk management needs, and ethical considerations. As the world grapples with challenges like climate change, inequality, and corporate governance, the role of banks in promoting sustainability and ethical practices will continue to grow in importance.

*Table 1 ESG Description*

| Item | Considerations | Examples | Actions |
| --- | --- | --- | --- |
| Environmental | Environmental considerations involve an organization's overall impact on the environment and the potential risks and opportunities it faces because of environmental issues, such as climate change and measures to protect natural resources, direct and indirect greenhouse gas emissions, management of toxic waste, and compliance with environmental regulations | • Climate change strategy<br>• Resource depletion<br>• Energy and water consumption and efficiency.<br>• Carbon footprint, including greenhouse gas emissions.<br>• Waste management.<br>• Air and water pollution.<br>• Biodiversity.<br>• Deforestation.<br>• Natural resource depletion. | • Publish a carbon or sustainability report<br>• Invest in renewable energy projects<br>• Provide financing for energy efficient buildings<br>• Limit harmful pollutants and chemicals<br>• Seek to lower greenhouse gas emissions and $CO_2$ footprint<br>• Reduces waste<br>• Offer green loans and sustainability-linked loans, e.g., electric vehicles, green home<br>• Install solar panels on bank branches<br>• Purchase and use renewable energy for bank operations<br>• Adopt a policy to minimize business air travel<br>• Set up e-waste recycling programs for electronic equipment<br>• Digitize banking services to reduce paper usage<br>• Commit to zero deforestation in lending and investment activities |
| Social | Social aspects address the company's relationships with internal and external stakeholders, e.g., employees, suppliers, customers, community | • Fair pay for employees, including a living wage.<br>• Diversity, equity and inclusion (DEI) programs.<br>• Employee experience and engagement. | • Operate an ethical supply chain<br>• Avoid overseas labor that may have questionable workplace safety or employ child labor<br>• Support and encourages diversity<br>• Have policies to protect against sexual misconduct<br>• Pay fair (living) wages |
| members and more | • Workplace health and safety.<br>• Data protection and privacy policies.<br>• Fair treatment of customers and suppliers.<br>• Customer satisfaction levels.<br>• Community relations, including the organization's connection to and impact on the local communities in which it operates.<br>• Funding of projects or institutions that help poor and underserved communities.<br>• Support for human rights and labor standards. | • Offer affordable housing loans to low-income groups<br>• Provide loans to minority-owned small businesses<br>• Set diversity targets for hiring and management<br>• Establish paid volunteer days for employees<br>• Donate a share of profits to local community organizations<br>• Offer financial literacy and education seminars<br>• Have supplier code of conduct covering labor practices<br>• Implement equitable parental leave and flexible work policies<br>• Audit algorithms for race or gender bias<br>• Hire ex-offenders and help their reintegration | |
| Governance | ESG governance standards examine how a company polices itself, focusing on internal controls and practices to maintain compliance with regulations, industry best practices and corporate policies | • Business ethics<br>• Company leadership and management.<br>• Board independence, diversity and structure.<br>• Executive compensation policies.<br>• Financial transparency and business integrity.<br>• Shareholder democracy & rights<br>• Regulatory compliance and risk management initiatives.<br>• Rules on corruption, bribery, conflicts of interest, and political donations<br>• Tax strategy<br>• Transparency & accuracy<br>• Whistleblower programs. | • Appoint a senior executive dedicated to ESG<br>• Establish ESG key performance indicators and executive incentives<br>• Report on ESG metrics and targets regularly<br>• Link executive pay to ESG performance<br>• Publish annual sustainability reports assurance<br>• Adopt a human rights policy statement<br>• Conduct ethical AI impact assessments<br>• Implement ESG risk analysis in lending decisions<br>• Train all employees annually on code of conduct<br>• Audit ESG disclosures annually<br>• Join external initiatives like the Equator Principles |

Although it is critical for the FIs, implementing ESG criteria is fraught with challenges, stemming from both internal and external factors. One of the primary challenges is data collection and quality. FIs often struggle to acquire reliable and standardized ESG data, especially when dealing with diverse geographical regions and industries. This inconsistency can lead to difficulties in accurately assessing and comparing the ESG performance of different investments or clients.

Another significant challenge is integrating ESG criteria into existing financial models and decision-making processes. Traditional banking models are primarily financially focused and may not adequately account for ESG risks and opportunities. Modifying these models to incorporate ESG factors requires not only technical adjustments but also a cultural shift within the organization. FIs must cultivate an understanding and appreciation of ESG issues among their staff and align these values with their overall business strategy.

Regulatory compliance and non-finalized "standards" present another hurdle. The ESG regulatory landscape is rapidly evolving and varies considerably across different jurisdictions. FIs need to stay abreast of these changes and ensure compliance, which can be resource-intensive and complex, especially for institutions operating internationally. The ESG industrial standards are generally under quick evolution to adapt the dynamics of the fast-changing world, e.g., the recent changes on Poseidon principle for shipping industry [8] and the SBTi and PCAF standard for facilitated emissions [9] [10].

Furthermore, there is the challenge of balancing financial objectives with ESG goals. While there is a growing recognition of the long-term financial benefits of ESG, FIs may face short-term trade-offs, such as foregoing potentially lucrative opportunities that do not meet ESG criteria. This balancing act requires a strategic approach and commitment to long-term sustainability goals.

Lastly, stakeholder engagement and communication are critical. FIs need to effectively communicate their ESG strategies and performance to a range of stakeholders, including investors, clients, and regulators. This requires transparent and consistent reporting, which can be challenging given the lack of standardized ESG reporting frameworks across all industrial sectors.

Therefore, the implementation of ESG criteria is a complex endeavor, challenged by issues related to data quality, integration into existing models, regulatory compliance, balancing financial and ESG objectives, and stakeholder communication. Addressing these challenges requires a comprehensive and strategic approach, underpinned by a strong commitment to sustainable banking practices [11]. FinTech (Financial Technology) thus plays a crucial role in this journey [1]. Among FinTech, a promising technology to prompt the implementation is Artificial Intelligence (AI) technologies. AI has been widely deployed in finance and economics, as discussed in the recent surveys, e.g., [12] [13] [14], where the corresponding challenges, techniques and opportunities of AI are listed. However, their focus in not for ESG, although some general principles are still applicable.

### 1.4 AI Impact and Opportunities

Given the complexity and critical importance of ESG criteria in banking, the application of AI technologies becomes not just advantageous but essential. AI's ability to analyze vast datasets efficiently and accurately is a key asset in managing the multifaceted aspects of ESG. For instance, in the environmental domain, AI can process and interpret large-scale environmental data, enabling FIs to assess climate-related risks with greater precision and foresight. This data-driven approach facilitates more informed decision-making in investment and lending, aligning financial activities with environmental sustainability [15]. Some daily activities, e.g., traded stocks, may require the assessment of additional ESG efforts, e.g., assess the structural balance sheet to check its effect on ESG score [11].

In terms of social and governance factors, AI can monitor and analyze patterns in a wide range of social and governance-related data, such as labor practices, supply chain management, and corporate governance structures. This capability allows FIs to identify and mitigate risks that may not be immediately visible through traditional analysis methods. Moreover, AI can help in detecting non-compliance with ESG standards, both within the bank and among its clients, by sifting through complex datasets for indicators of potential issues.

Furthermore, AI enhances customer engagement and personalization in the context of ESG. By analyzing customer data, FIs can offer tailored ESG-related products and services, meeting the growing demand for sustainable and ethical banking options.

In essence, AI empowers FIs to navigate the ESG landscape with greater accuracy, efficiency, and foresight. It enables a more nuanced understanding of ESG risks and opportunities, thereby fostering more sustainable and responsible banking practices. In an era where ESG considerations are increasingly integral to the financial sector's viability and success, leveraging AI becomes a strategic necessity. Next sector will provide more details.

## 2. AI Applications in ESG

AI is being used to implement or improve ESG in many aspects [1] [16]. Some main directions are categorized in Table 2. For easy reference, we further classify the applications into 4 classes: U, E, S, G, where U is for the universal applications in all ESG domains, while E, S and G indicate environmental, social and governance, respectively. They are listed in Table 3, Table 4, Table 5 and Table 6 separately. Some applications will be further discussed in the later sections as part of a wider or bigger process.

*Table 2 AI application category*

| Category | Description |
|---|---|
| Data collection & analysis | - Use natural language processing (NLP) to extract unstructured ESG data from various sources like financial reports, news, filings.<br>- Apply machine learning (ML) or deep learning (DL) algorithms to analyze large ESG datasets to identify trends, correlations and insights. For example, correlating carbon emissions data with transportation fleets and routes.<br>- Develop predictive models using time series analysis to forecast future ESG metrics like water usage, waste generation etc. This supports target setting.<br>- Use sentiment analysis on social media conversations and stakeholder reviews to measure perceptions of the company's ESG reputation. |
| Risk assessment | - Build ML models to analyze economic, geopolitical and environmental data to quantify climate change risks like sea level rise, droughts affecting operations.<br>- Use scenario analysis to simulate different risk events like supply chain disruptions from extreme weather and estimate potential financial impacts.<br>- Apply NLP to reviews, complaints to identify emerging human rights violations, ethical issues in the company or supply chain.<br>- Improve their lending practices. For example, AI can be used to assess a borrower's creditworthiness more accurately, which can help to reduce the risk of lending to borrowers who are unable to repay their loans. |
| Investment | - Use quantitative algorithms to screen potential investments for material ESG risks as well as alignment with sustainability goals.<br>- Optimize ESG investment portfolios to maximize sustainability objectives while managing financial risk using techniques like reinforcement learning. |
| | - Evaluate companies' long-term value and investment potential by analyzing ESG performance such as carbon efficiency. |
| | - Identify sustainable investment opportunities. For example, AI can be used to screen companies for ESG criteria, such as having low carbon emissions or strong labor practices. |
| Compliance | - Automate regulatory filings by using NLP to analyze guidelines and produce compliant reports. |
| | - Monitor regulations worldwide using ML to flag relevant new requirements, e.g., monitor financial markets for ESG risks and identify companies that are violating ESG regulations. |
| | - Use robotic process automation (RPA) to gather and validate ESG data for compliance. |
| Engagement | - Analyze stakeholder feedback and sentiments using NLP to identify areas for improving ESG engagement. |
| | - Generate customized communications to stakeholders about sustainability initiatives using natural language generation. |
| | - Build chatbots to interact with customers and gather direct feedback on ESG topics |

*Table 3 Universal applications*

| Index | Item | Description | Comments |
|---|---|---|---|
| U.1 | ChatBot: Frequent Questions and Answers (FAQ) | For given FAQ datasets (e.g., AQ pairs in csv/excel formats), use AI technique (e.g., similarity search based on text embedding vectors) to find the most suitable answer to the users' questions. If not found, return null. | Available & matured solution for text; other formats may be achievable via engineering effort |
| U.2 | ChatBot: Document-based/General QA | For given knowledge datasets (e.g., various materials in pdf/word/excel/ppt formats), use AI technique (e.g., retrieval augmented generation (RAG) together with LLM) to generate suitable answers to the users' questions. If not found, return null. Ideally, it shall support Omni-channel, and integrated with CRM. Open-loop chatbot is generally not suitable for FIs. | Available & matured solution for text; other formats may be achievable via engineering effort |
| U.3 | ESG data collection, e.g., Information extraction from documents | Use NLP and text analytics to extract ESG data from annual reports, sustainability reports, websites etc. Extract the useful values from the given text documents, e.g., the net zero target, GHG emission value, etc. | Available & matured solution for general text; the accuracy may be slightly lower when processing multiple tables (unstructured) |
| U.4 | ESG reporting automation | Use NLP and data analytics to reduce reporting time and errors for sustainability reports. AI-driven automation can streamline the reporting process, reducing manual effort and human errors. This can lead to more consistent and timely sustainability disclosures. Accurate reporting of ESG data brings in the needed transparency for investors and gives them an understanding of a company's ESG direction and progression. | Relatively matured for fixed format; manual review is required for "free-style" content by generative AI |
| U.5 | ESG product development and assessment | Use ML to identify market needs for new sustainable banking products. For example, design products with lower environmental impact using AI-driven simulations. Nike uses AI to design more sustainable athletic shoes. | |
| U.6 | ESG competitive benchmarking | Comparing against peers' ESG performance using internal and external data and analytics. The similar method can be applied to the company for its own operational performance | |
| U.7 | ESG scenario analysis | Model ESG factors into stress testing scenarios to gauge strategic resilience. AI can simulate various scenarios to assess the potential impact of different decisions on a company's sustainability performance. This can aid in disclosing information about the company's resilience and adaptability. | |
| U.8 | ESG trending monitoring | AI tools analyze the news feeds, company filing, third-party aggregated data and other available information to monitor ESG related themes. | Mainly use NLP techniques to extract the relevant information |

*Table 4 Environmental applications*

| Index | Item | Description | Comments |
|---|---|---|---|
| E.1 | Environmental metrics from satellite and sensor data | Collects satellite and sensor data to determine environmental impact and physical risk exposures. For example, with its extensive geographic coverage, this data is instrumental in confirming companies' carbon emissions and examining their effects on various ecosystems, including air pollution, waste generation, deforestation, and flooding, among others. Additionally, this kind of data plays a crucial role in climate risk stress testing models. The insights derived from these models have been notably enlightening, in our perspective. [17] IoT techniques are required to monitor the comprehensive GHG data. | Stallite imaging processing can detect multiple GHG, although not all. Similar techniques can be applied to analyze social activites to estimate poverty and economic well-being. |
| E.2 | Carbon footprint measurement | Use AI to analyze data across operations and quantify scope 1, 2 and 3 GHG emissions. This provides insights for setting emission reduction targets. | |
| E.3 | Renewable energy trading | Develop algorithms to trade carbon credits and renewable energy certificates to offset emissions. | |
| E.4 | Smart offices and Green building design/ implementation | Use IoT sensors, computer vision and AI to automate lighting, HVAC systems optimization for energy efficiency and cost savings. Utilize generative design and AI to create optimal energy efficient branch layouts | |
| E.5 | Paperless / digital banking | Digitize banking processes, e.g., account opening and loan processing, and implement AI-OCR and NLP for auto-categorization and extraction to reduce paper usage. | OCR has high accuracy for printed text, while still needs to be tuned for specified and/or mixed handwriting texts in |

| Index | Item | Description | Comments |
|-------|------|-------------|----------|
| | | | multilangual environments. |
| E.6 | Sustainability loan pricing | An automated, AI-driven system offers FIs the capability to assess an enterprise's ESG key performance indicators (KPIs) and associated pricing fluctuations efficiently. This system is designed to aggregate sustainability data, conduct analyses, and deliver pertinent insights to the stakeholders involved. | |
| E.7 | Energy usage analytics and management. | Apply IoT sensors and AI models to track water/electricity consumption patterns across bank locations. This identifies opportunities for efficiency improvements and predictive maintenance of crucial infrastructure and prevention of accidents and disruptions. An example is to Implement AI-controlled HVAC systems to reduce energy usage in buildings. | It is essentially a resource optimization problem, in particular, useful for IT center |
| E.8 | ESG environmental forecasting | Predict future ESG metrics like emissions, water usage based on historical data. | |
| E.9 | Supply Chain Sustainability | Trace product origins and assess suppliers' sustainability using AI-powered blockchain. For example, Walmart uses blockchain to trace the origin of its produce. Smart contract might also be helpful.<br>Support circular supply-chain with recovery and recycling. Use sharing platform and product as a service concept. Extend product life cycle. | It is also applicable for optimized logistics and scheduling to improve transportation sustainability. |
| E.10 | Carbon Offsetting | Identify and invest in carbon offset projects using AI-driven analysis. For example, Shell uses AI to identify and invest in forestry projects for carbon offsetting. | |
| E.11 | ESG climate risk analysis | Model climate change impacts on credit risk exposure, operational risks. | |

*Table 5 Social applications*

| Index | Item | Description | Comments |
|-------|------|-------------|----------|
| S.1 | Diversity and inclusion analytics | Apply NLP and ML on client or employee engagement surveys, HR complaints, exit interviews to detect biases and barriers.<br>For example, Analyse pay data to address gender pay disparities. Salesforce used AI to identify and close its gender pay gap. | In implementing, e.g., identify gender of clients of SMEs to understand the weights of females |
| S.2 | Ethical AI audits | Perform automated audits of AI systems using testing datasets to detect algorithmic biases and ensure responsible AI. | |
| S.3 | Microfinance credit underwriting | Use alternative data and ML models to evaluate creditworthiness of unbanked or thin-file applicants. Expand financial inclusion. | |

| S.4 | ESG data transparency | Adopt blockchain, distributed ledger technology to record ESG initiatives' impacts and enable traceable disclosures. | In implementing, e.g., CarbonMarket |
|---|---|---|---|
| S.5 | Financial inclusion | Use alternative credit risk models leveraging ESG data to broaden access. | |
| S.6 | Sustainable procurement | Incorporate ESG criteria into supplier selection algorithms. AI can also track vendor ESG performance over time. | |
| S.7 | ESG sustainability bonds | Automate analysis of use of proceeds from sustainability bonds issuance. | |
| S.8 | ESG cybersecurity | Leverage AI to bolster ESG data privacy, security protections. | |
| S.9 | ESG employee training | Deliver personalized ESG e-learning based on roles and knowledge gaps. | Content generation is rather easy now based on the latest multi-modal tools |
| S.10 | ESG employee sentiment | Measure employee sentiments on internal ESG efforts through surveys and NLP. | |
| S.11 | ESG customer engagement | Use chatbots with personalized recommendations to promote green financial products to customers. | |
| S.12 | ESG customer analysis | Analyze customer data to identify preferences and behaviors regarding ESG products. | |
| S.13 | ESG portfolio optimization | Use quantitative models to construct optimal portfolios balancing ESG factors, risk and return. | |
| S.14 | ESG sustainability investment recommendation | Build automated solutions to provide retail investors personalized ESG investment recommendations and portfolios. | The latest LLM links more contents to provides more insights for recommendations. |
| S.15 | Community Engagement | Analyse community feedback to inform sustainability initiatives. For example, Mining companies use AI to gauge community sentiment and adjust their practices accordingly. | |
| S.16 | ESG investor relations | Automated analysis of investors' ESG priorities to guide engagement. | |
| S.17 | ESG marketing targeting | Micro-segmenting and predictive modeling to optimize ESG marketing campaigns. | |
| S.18 | ESG customer lifetime value | Incorporate ESG criteria into customer lifetime value models. | |
| S.19 | Shakeholder management | Within the domain of corporate responsibility and sustainability, the management of stakeholders— | The relation is generally complex as the |

encompassing owners (inclusive of investors), employees, governments, suppliers, customers, NGOs, and the media—perhaps holds greater significance than numerous other considerations that companies must address. Utilizing a graph-based methodology could prove effective in delineating these relationships and in overseeing their respective interests.

connections are always multiple directional.

*Table 6 Governance applications*

| Index | Item | Description | Comments |
|-------|------|-------------|----------|
| G.1 | ESG scoring/rating/ indexing/ bench- marking | Develop algorithms to assess and quantify the bank's overall ESG performance. Create specialized indices to track securities meeting ESG criteria. Compare ESG performance against industry peers using AI analytics. for example, Sustainalytics provides AI-driven ESG benchmarking services. | |
| G.2 | ESG compliance and fraud detection, e.g., ESG green/social washing & tax eva- sion screening | Build automated tools to screen potential investments for ESG risks and criteria compliance, e.g., analyzing transactions for ESG-related misrepresentations or fraudulent activities. Anti-Money Laundering (AML) Compliance is important. AI may also reduce the risk of green washing or social washing – the practice of making a company appear more ESG-friendly than it really is. For example, AI can help the agency exclude company statements that mention sustainability practices, which are not material to the business and are unlikely to matter to investors. | |
| G.3 | ESG auditing | Applye intelligent process automation for internal and external ESG audits. | |
| G.4 | ESG board intelligence | Analyze board materials to provide insights into ESG governance. | It covers multiple aspects mentioned earlier, e.g., report automation, ESG auditing, etc. |
| G.5 | Supply chain transparency | AI can help trace the environmental and social impact of products and materials across the supply chain. This transparency can lead to more accurate and detailed disclosures about a company's supply chain practices, e.g., steering clear of the businesses with people working in slavery-like conditions. | |
| G.6 | ESG accounting | Automating measurement and accounting of ESG initiatives' impacts and costs. | |
| G.7 | ESG due diligence & materiality analysis | Conduct AI-driven due diligence on ESG issues during mergers and acquisitions. For example, Law firms use AI for ESG due diligence in legal transactions. | It's essentially part of the compliance process. |
| | | AI can assist in identifying the most relevant ESG issues for a company based on data analysis and stakeholder engagement. This ensures that disclosures focus on the topics that matter most to stakeholders. | |
| G.8 | ESG controversies monitoring | Track ESG-related incidents and social media to assess reputation risks. Provides textual analysis to measure companies' ESG incidents and commitments. [18][2] Textual analysis can identify companies' controversies and important ESG news. ESG data providers (e.g. RepRisk and Truvalue Labs) can use NLP tools to analyse real-time company information to measure controversies surrounding environmental policies, working conditions, child labour, corruption, etc. For example, RepRisk analyses more than 80,000 media, stakeholders and third party sources daily, detecting incidents that occur in companies' ESG policies. This type of analysis can be very informative, adding value to ESG investment processes. | Fake news identification is another aspect to be considered to avoid data pollution. |
| G.9 | ESG compliance monitoring | Scan regulations, filings, press releases for new ESG-related compliance requirements. | |
| G.10 | ESG risk modeling | Develop risk models by incorporating ESG factors for credit underwriting, lending decisions. | |

### 2.2 Industrial Practices

AI's role in ESG is transformative, offering industries innovative tools to manage and improve their sustainability practices, engage with stakeholders more effectively, and make data-driven decisions that align with their ESG objectives. Many financial institutions have applied AI into the ESG-related works. Table 7 listed some typical industrial examples. For example, Standard Chartered Bank applied the capability of LLM to extract the climate risk information from clients' annual reports and ESG reports and answer the risk questionaries to accelerate the assessment process.

*Table 7 FIs' AI practices*

| Case provider | Case description |
|---|---|
| Bank of America | applies AI in its risk management principles, focusing on ESG factors and optimizing efficiencies in business practices |
| Citibank | exploring AI applications in business and finance, including the integration of AI in ESG strategies |
| HDFC Bank | tapping into AI for customer operations and loan automation, which could have implications for their ESG strategies. |
| HSBC | launched an AI-powered index that tracks companies benefiting from improving ESG risk. This index tracks the performance of over 1,000 liquid stocks of global companies, with ESG risk measured by calculating a score |
| JP Morgan Chase | collaborates with Datamaran to integrate data-driven and dynamic double materiality into its ESG integration process, utilizing AI-driven technology |
| Morgan Stanley | uses AI applications for sustainable investing, including improving the accuracy of ESG metrics and employing AI-powered satellite imaging |
| NatWest | launched an AI solution to enhance ESG data for SMEs, linking individual customers and properties with relevant ESG data |
| OCBC | implementing AI technologies, such as generative AI chatbots, to enhance employee productivity and support their ESG initiatives |
| Standard Chartered Bank | leveraging AI tools to organize ESG data more effectively, as part of their commitment to sustainable practices. Applied NLP/LLM in climate risk assesment. |
| Wells Fargo | exploring AI applications to enhance customer service and other banking operations |

We might also analyze the winner and finalist projects of Earthshot prize in 2023 to understand how the AI can help on these projects [19]. See Table 8 for the details.

*Table 8 AI for Earthshot projects*

| Project Category | Project examples | Potential AI Applications |
|---|---|---|
| Forest Restoration Projects | Acción Andina, Belterra, Freetown the Treetown | - Satellite imagery analysis for deforestation detection.<br>- Predictive modeling for future deforestation risks.<br>- Targeted reforestation effort optimization. |
| Soil and Air Quality Initiatives | Boomitra, GRST, ENSO | - Soil data analysis for health and carbon sequestration insights.<br>- Air quality data processing for pollution source identification and pattern analysis. |
| Marine Conservation Efforts | WildAid Marine Program, ABALOBI, Coastal 500 | - AI-driven image recognition for marine biodiversity monitoring.<br>- Predictive modeling for marine ecosystem health and human impact assessment. |
| Waste Management and Recycling | S4S Technologies, Circ, Colorifix | - Automation in waste sorting processes.<br>- Optimization of recycling methods.<br>- Supply chain efficiency improvements in waste management. |
| Water Treatment | Aquacycl | - Water demand prediction.<br>- Optimization of treatment processes.<br>- Anticipatory maintenance in wastewater facilities. |
| Sustainable Fisheries | ABALOBI | - Data analysis for sustainable fishing practices.<br>- Ocean condition monitoring.<br>- Fish population dynamics prediction. |
| Urban Planning for Air Quality | Polish Smog Alert, ENSO | - Urban pollution modeling.<br>- Traffic and industrial emission management strategies. |

*Figure 1 Interaction between AI and ESG world*

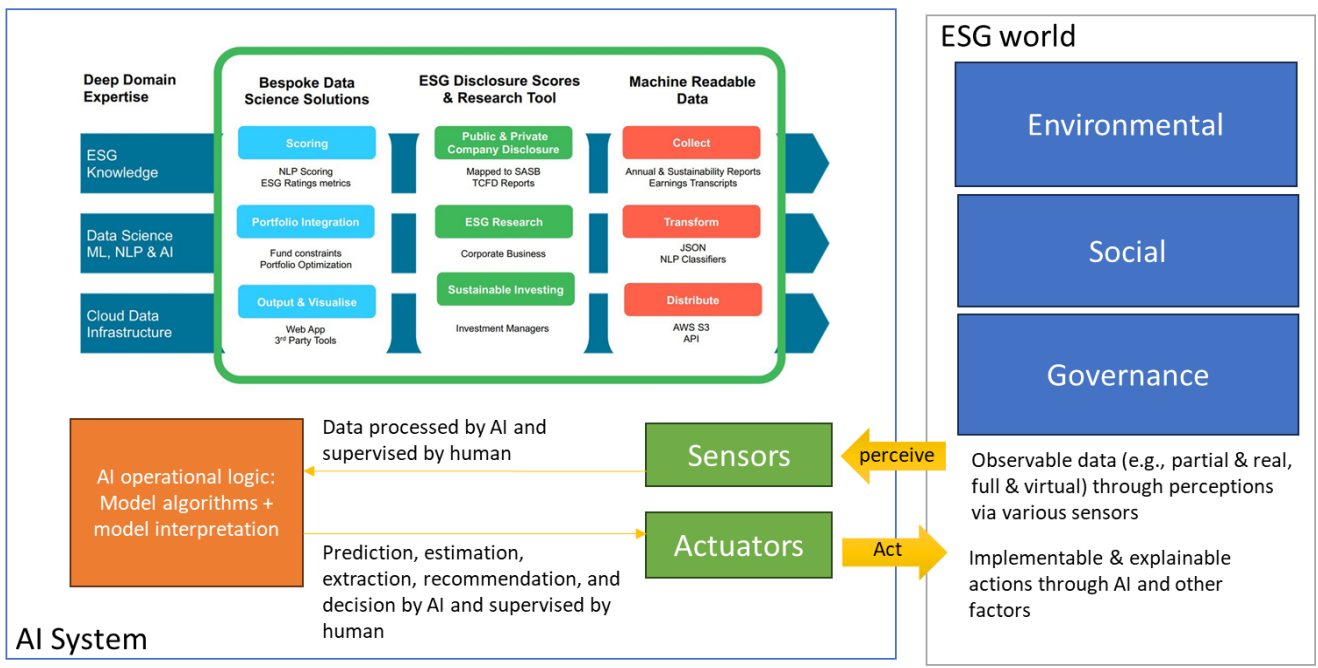

## 3. Data Matters

### 3.1 IT and Data Infrastructure

In order to fulfill these rising expectations on ESG adoption and integration, FIs are required to modify their information technology (IT) systems and Data Infrastructure for the systematic collection, aggregation, and dissemination of a wide spectrum of ESG data. The core components of contemporary data architecture encompass data integration, engineering, quality, and observability, along with a data catalog (asset management). It also involves data governance and privacy, integration of APIs and applications, a data marketplace, various data products, and data mastering. Presently, numerous financial institutions lack a holistic strategy for embedding ESG data into their existing frameworks.

Achieving this objective necessitates substantial modifications to IT infrastructure, encompassing aspects such as applications, data integration, architecture, and governance. These new applications are not limited to the management and acquisition of ESG data but also extend to models for financed emissions, physical emission, climate risk, ESG scorecards, climate-related stress tests, and climate-adjusted ratings [20]. It is imperative for ESG data to be integrated into existing procedures, like credit approval and decision-making processes. Furthermore, FIs are required to revamp their data architecture, establish a strategy for data collection, and restructure their data governance model to effectively manage and report ESG data.

By prioritizing the right investments from the outset, IT leaders in the financial sector can swiftly develop these new capabilities and solutions into a unified ESG platform, thereby avoiding the accumulation of technical debt [21].

Defining potential ESG platform solutions involves a multifaceted approach that prioritizes data integration, accessibility, and modernization. Firstly, establishing a central data platform, seamlessly integrated with existing finance and risk platforms, is crucial. This creates a unified source of truth, enhancing data accuracy and consistency. Developing a robust data model to capture ESG data at the certificate level is another key aspect. This model should facilitate integration with third-party data providers through APIs, ensuring compliance with ESG data policies.

In addition to these measures, the platform should enable investors to have real-time visibility into the ESG attributes of their investment portfolios. This feature empowers investors to make more informed, sustainable decisions. Moreover, transitioning from legacy ESG solutions to a cloud-based framework is important for future scalability. This shift not only reduces technical debt but also modernizes the technological infrastructure, paving the way for future advancements.

Expanding further, potential ESG platform solutions should also include advanced analytics capabilities for predictive modeling and trend analysis in ESG metrics. Automated reporting tools for generating standardized and customized ESG reports would facilitate compliance and stakeholder communication. Furthermore, incorporating AI and ML algorithms can enhance the processing and interpretation of large-scale ESG data, improving decision-making processes. Lastly, ensuring robust cybersecurity measures within these platforms is vital to protect sensitive ESG data and maintain investor trust.

This potential ESG platform solutions should generally focus on data integration, real-time visibility, technological modernization, advanced analytics, automated reporting, AI-enhanced data processing, and stringent cybersecurity, all cohesively working to drive sustainable investment practices.

At the same time, this shift has brought about a range of risks associated, e.g., digitalization. Current focus areas, like the United Nations Sustainable Development Goals (SDGs) and established ESG frameworks, tend to emphasize climate-related issues, often overlooking digital concerns. This oversight has permitted companies that may negatively impact the digital environment to skirt their ESG obligations without substantial repercussions to their investments or reputations. Integrating Digital-ESG into sustainability strategies offers a way to balance the benefits of technological advancements with the protection of citizens, both in digital and physical realms, as well as the broader global society [22].

### 3.2 Data Collection, Cleaning and Aggregation

Challenges related to data and information are significant hurdles in ESG analysis. As ESG factors increasingly influence investment decisions, the demand for high-quality, real-time ESG analysis escalates. However, analysts and investors face several data-related obstacles historically, e.g.,

- The presence of unstructured or incomplete data or inaccurate data.
- The prevalence of qualitative and more ambiguous forms of information.
- Delays in the flow of information.
- Evolving reporting practices among companies.
- Variability in reporting templates and structures.

In the realm of Sustainable Finance, the transition from a historical data scarcity to a current overflow of information presents a paradoxical challenge. This abundance is often not standardized, compromising data reliability and utility. The pursuit of high-quality data has been a key focus, with recent advances indicating promising solutions.

Evaluating a company's sustainability requires collecting ESG data from various sources such as open-source publications, internal client documents, or external providers. Some companies and organizations provide data aggregation services, as listed in Table 9. Once gathered, this data should be stored on an easily accessible platform, where AI plays a crucial role in unifying diverse data sets for specific client analysis. The data ops pipeline is also important. These challenges complicate the process of conducting meaningful, accurate, and timely ESG analysis.

*Table 9 ESG data providers*

| Provider | Description | Website |
|---|---|---|
| Bloomberg | Provide global coverage of ESG data; mainly on listed or public companies | https://fin-science.com/en/blog/esg/top-5-esg-data-providers-rating-and-report/ |
| BvD (Moody's) | Provide ESG ratings, analytics, sustainability ratings, and sustainable finance reviewer/certifier services using data from Moody's. The BvD Orbis users can now add RepRisk's ESG risk metrics to the already substantial arsenal of company information available to you through our flagship global database of more than 200 million private companies | https://www.bvdinfo.com/en-gb/blog/compliance-and-financial-crime/product-update-using-orbis-to-assess-environmental-social-and-governance-esg-risk |
| CDP | CDP (formerly the Carbon Disclosure Project) ESG Rating is a unique rating that identifies the best ESG-integrated investment funds, based on their ESG performance | https://www.cdp.net/en/scores/cdp-scores-explained |
| FTSE Russell / LSEG | Provide ESG ratings using a company's Theme Exposure and Theme level score assessment to calculate range of assessments that allow investors to understand a company's ESG practices in multiple dimensions. | https://www.ftserussell.com/products/indices/esg |
| ISS | Provide ESG data and research on companies. The system's goal is to take the ESGY Scorecard directly to the industry level. | https://www.issgovernance.com/esg/ |
| MSCI | Offer data on ESG. Its ratings aim to measure a company's management of financially relevant ESG risks and opportunities. | https://www.msci.com/our-solutions/esg-investing/esg-ratings |
| Refinitiv / LSEG | Provides comprehensive ESG data. Their ESG scores are designed to transparently and objectively measure a company's relative ESG performance, commitment and effectiveness across 10 main themes (emissions, environmental product innovation, diversity and inclusion, human rights, shareholders, etc.) based on publicly-reported data | https://www.refinitiv.com/content/dam/marketing/en_us/documents/methodology/refinitiv-esg-scores-methodology.pdf |
| Sustainalytics | An independent ESG research and rating agency that provides data and research on the environmental and social performance of companies. | https://connect.sustainalytics.com/hubfs/SFS/Sustainalytics%20ESG%20Risk%20Rating%20-%20FAQs%20for%20Corporations.pdf |
| S&P Global | Provide access to transparently disclosed ESG data points for companies assessed in the S&P | https://www.spglobal.com/esg/solutions/esg-data-intelligence |

### 3.3 Data Governance Model

Building a robust ESG data governance model for AI applications necessitates a comprehensive and coordinated approach, incorporating several key actions:

o Developing a comprehensive ESG data taxonomy to standardize data classification and ensure consistency in data collection, storage, and analysis. In other words, an ESG data catalog [23] is an essential step for data governance. Some organizations also called it as Master Data management (MDM) [20].

o   Assigning central ownership and responsibility within the organization is fundamental. This can be achieved by appointing a dedicated ESG data officer, who acts as the central point of contact and ensures cohesive ESG data management across the organization.

o   Establishing a cross-functional steering committee for ESG data governance is critical. This committee should include leaders from business, technology, data, risk, and finance departments, fostering joint accountability and streamlined decision-making processes.

o   Implementing ESG data controls is essential to maintain compliance with various regulatory frameworks. This involves setting up mechanisms to verify and track compliance markers, such as the assignment of certificates to investments. Conducting regular audits and reviews of the ESG data governance framework is required to ensure its effectiveness, relevance, and compliance with evolving ESG standards and best practices.

o   The ESG data governance framework must be adaptable to changes in market demand and region-specific regulatory requirements. For instance, it should be capable of accommodating investments in emerging sectors like offshore wind turbines and aligning with local regulations, such as the prospective bans on investments in combustion engines in specific countries.

o   Integrating advanced analytics (e.g., AI/ML) and reporting tools to enhance the analysis, interpretation, and communication of ESG data, facilitating better decision-making and stakeholder reporting.

### 3.4 Data Assets

The integration of AI into ESG data asset creation, management, and applications marks a significant advancement in the way businesses and organizations approach sustainability and ethical governance. AI's ability to process and analyze large volumes of data from varied sources is particularly beneficial in the complex and evolving field of ESG, where data is often unstructured and dispersed across multiple platforms. [24]

In the creation of ESG data assets, AI algorithms are adept at collecting and synthesizing information from a wide array of sources, including satellite imagery, environmental sensors, social media, news outlets, and corporate sustainability reports, as discussed in the last sector. This comprehensive data collection enables the creation of rich, multi-dimensional ESG data assets. AI's advanced analytics can identify patterns and correlations within this data, uncovering insights that might be missed by human analysts. For instance, ML models can predict environmental impacts based on current trends, or NLP can analyze sentiment in social media to gauge public opinion on corporate social responsibility practices.

When it comes to the management of ESG data assets, AI offers unparalleled efficiency and accuracy. AI systems can continuously monitor and update ESG data in real-time, ensuring that the information remains current and relevant. This is particularly important in a rapidly changing world where environmental conditions, social dynamics, and governance regulations can shift quickly. AI-driven systems can also automate the reporting process, generating comprehensive and compliant reports that meet the standards of regulatory bodies and stakeholders. This automation not only saves time and resources but also reduces the likelihood of human error, enhancing the reliability of ESG reporting.

By utilizing the ESG assets, AI has the potential to revolutionize how companies implement and track their sustainability and governance initiatives. Predictive analytics, a key feature of AI, can forecast future ESG trends, allowing companies to proactively address potential risks and capitalize on emerging opportunities. For example, AI models can predict the impact of climate change on supply chains, enabling companies to make informed decisions about resource allocation and risk management. Additionally, AI can tailor ESG strategies to specific industries and companies, taking into account unique environmental impacts, social responsibilities, and governance challenges. This customization ensures that ESG initiatives are not only effective but also aligned with the company's specific goals and values.

Moreover, AI enhances stakeholder engagement in ESG efforts. By analyzing stakeholder data, AI can help companies understand the concerns and priorities of their customers, employees, and investors. This understanding enables businesses to develop ESG initiatives that resonate with their stakeholders, fostering a sense of trust and commitment. AI-driven platforms can also facilitate transparent and interactive communication with stakeholders, allowing companies to share their ESG progress and receive feedback in real-time.

## 4. Model Matters

Many types of AI models have been applied in the industry, from traditional expert systems, general ML and deep learning models to the latest generative-AI based on large language models (LLMs) [25] [26]. There are many considerations, e.g., application-specific issues, different between traditional and AI-powered solutions, responsible AI (see Section 5), sustainable AI (see Section 6). Below, we provide some examples to discuss the details. The first one is the ESG scoring model using AI. The other two consider Net Zero model and Greenwashing model, respectively.

### 4.1 ESG Metrics, Rating and Scoring

One of the important insights from the ESG data assets using statistical and AI models is the ESG rating and scoring. These metrics evaluate and rate companies based on their performance in ESG aspects with certain criteria, e.g., a company's carbon footprint, labor practices, corporate governance structures, and community engagement. This evaluation helps investors and stakeholders make informed decisions regarding the sustainability and ethical practices of these companies. AI technologies are instrumental in developing more precise ESG rankings for companies. The infographic illustrates a typical ESG methodology, encompassing several stages:

1.  The process initiates with the identification and creation of an ESG metrics dataset. This dataset is compiled through web scraping and other methods, drawing from a variety of sources as discussed in the last section. These include private data from internal records, public data from government and non-governmental organizations, corporate websites and reports, and third-party information predominantly based on sustainability reports.

2.  The data thus gathered undergoes a Quality Assurance (QA) phase, after which it is processed through ML and NLP techniques. Some typical techniques comprise:

- Word embeddings: This involves mapping words to numerical representations that encapsulate their meanings, semantic relationships, and the various contexts in which they are used.

- Topic and theme tagging: Assigning relevant topics and themes to the data for enhanced processing.

- Sentiment analysis: This step involves categorizing sentiments as negative, positive, or neutral, and aggregating these sentiments at the topic and theme level.

3. Subsequently, the processed data is standardized into a scoring system, ranging, for example, from 1 to 100 (regression model), or categorized into indices such as a 1-5 distribution (classification model), with the high value representing the highest risk or poorest-performing companies, and the low value indicating the lowest risk or best performers. Topic weighting is also implemented, enhancing the correlation of ESG scores with financial performance. The importance of topics varies by sector. For example, in a credit card company's ESG assessment, topics like energy management hold minimal relevance compared to data privacy. Conversely, for a utility company, energy management is more pertinent than data privacy.

4. Finally, the results from these models, constituting the ESG rankings, are integrated into an overall company resiliency ranking.

*Table 10 Comparison of traditional and AI-power rating*

| Aspect | Traditional ESG Rating Agencies | AI-Powered ESG Rating Organizations |
|---|---|---|
| Data Processing and Analysis | Rely on manual data collection and analysis, subject to human bias, e.g., analyst. | Use algorithms, ML, and NLP to analyze large data sets efficiently, reducing human bias. |
| Speed and Scalability | Slower due to manual processes; limited in the number of companies assessed. Usually, update semiannaully or annually | Faster updates and ability to scale up to assess more companies due to automation. Typically, update in daily or weekly. |
| Consistency and Objectivity | Potential for inconsistencies and subjective interpretations. | More consistent and objective, though not immune to underlying data biases. |
| Predictive Insights | Focus on current and past performance based on available data. | Capable of offering predictive insights about future ESG performance and risks. |
| Customization and Flexibility | Limited customization and flexibility due to manual processes. | Greater customization and flexibility, adapting quickly to new data sources or changing ESG criteria. |

*Figure 2 Stages of Ranking & Scoring*

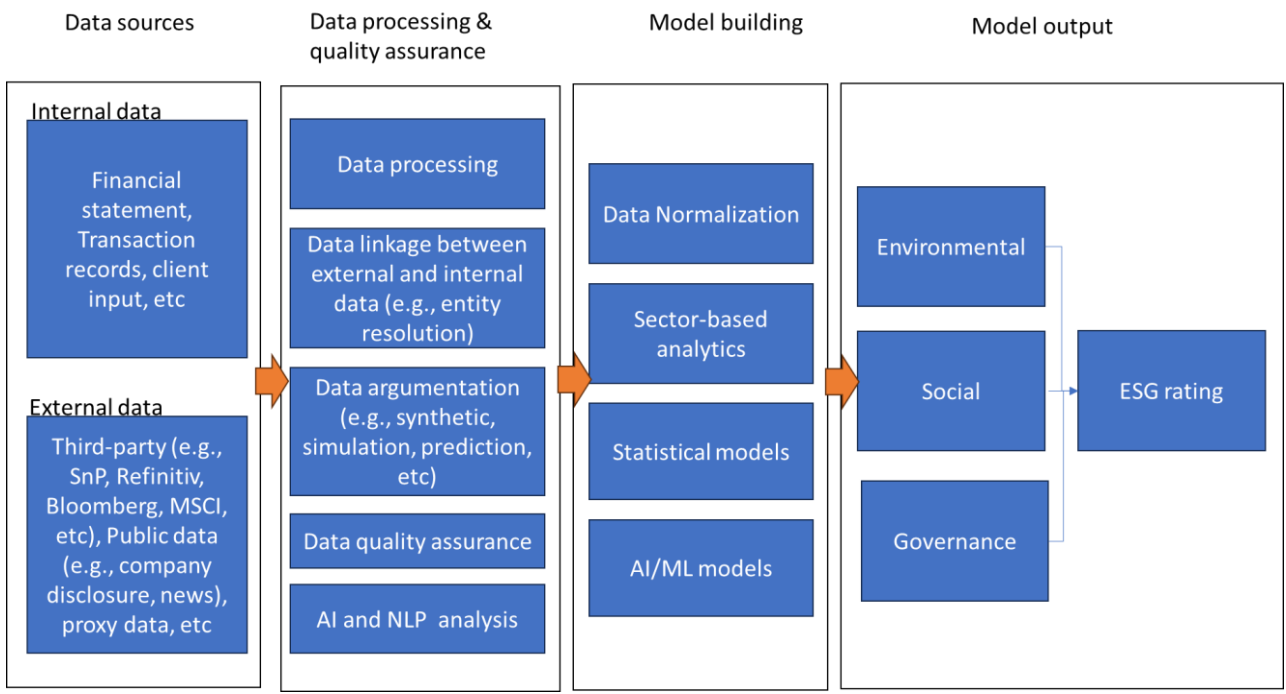

| Data sources | Data processing & quality assurance | Model building | Model output |

As shown in Figure 2, the whole process can be AI-driven. Some examples of AI-driven ESG rating providers include:

- RepRisk: According to their methodology page, RepRisk conducts daily scans of over 100,000 public sources and stakeholders in 23 languages. As of 2022, their dataset encompasses over 225,000 companies linked to risk, with approximately 4% being publicly listed and about 93% being private entities. [27]

- Truvalue Labs: Based on their methodology document, Truvalue Labs processes information from upwards of 100,000 sources. They offer four distinct scores to aid a wide array of investment strategies: Insight, Pulse, Momentum, and Volume scores, which form the bedrock of all Truvalue products. [28]

- ESG Book: Originating from Arabesque in 2018, ESG Book merges advanced technology with unique research. Their ESG Data Solutions facilitate analysis at both the company and portfolio levels, providing ESG data for over 25,000 companies. [29]

*4.2 Net zero model*

Net zero refers to achieving a balance between the amount of greenhouse gases emitted into the atmosphere and the amount removed from it. This concept is central to the strategies aimed at mitigating climate change, where the goal is to ensure that any emissions are counterbalanced by absorbing an equivalent amount from the atmosphere, effectively reducing the net emissions to zero [30].

The common behaviors associated with achieving net zero are multifaceted and interdependent. Firstly, there is a concerted effort towards emission reduction, which involves actively diminishing greenhouse gas emissions through enhanced energy efficiency, the adoption of sustainable practices, and the implementation of cleaner production methods. Alongside this, carbon capture and storage play a crucial role, wherein technologies are employed to extract $CO_2$ from the atmosphere and store it securely, mitigating its environmental impact. Another pivotal behavior is the integration of renewable energy sources, such as solar and wind power, which represents a shift away from traditional fossil fuels towards more sustainable energy solutions. Furthermore, sustainable supply chain management is essential, ensuring that every stage of the supply chain adheres to environmentally friendly practices. Finally, the regular monitoring and reporting of greenhouse gas emissions are fundamental behaviors, as they provide essential data for assessing progress and making informed decisions towards achieving net zero goals. These behaviors collectively form the backbone of strategies aimed at balancing greenhouse gas emissions with their removal from the atmosphere.

The general way for net zero initiative in FIs is to set up a target first, build the corresponding net zero models for various industries (usually include the baseline of emission, and reference and projection) and finally embed the models into the financial forecasting or other decision-marking processes [31]. However, there are many challenges in the building the net zero models. Table 11 addresses those challenges and their potential AI solutions. Each solution is tailored to address the specific aspects of the challenge, leveraging the strengths of AI in data processing, predictive analytics, and optimization to enhance the modeling of net zero strategies.

*Table 11 AI solution for Net Zero model*

| Challenges | Description | Potential AI Solution |
|---|---|---|
| Data Complexity and Volume | Handling vast and complex data related to emissions across various industries. | Use of big data analytics and ML to process and analyze large datasets efficiently. |
| Dynamic Nature of Emissions | Emissions vary over time and by industry, requiring adaptable models. Different scopes may need different approaches. | Development of flexible AI models that can adapt to varying emission patterns and industry-specific factors. |
| Integration of Diverse Data Sources | Combining data from different sources and sectors for comprehensive analysis. | Implementing AI algorithms capable of integrating and synthesizing diverse data streams for holistic insights. |
| Predictive Accuracy | Accurately predicting future emissions and the effectiveness of mitigation strategies. | Employing predictive analytics and advanced ML techniques for precise forecasting of emissions and strategy outcomes. |
| Balancing Economic and Environmental Goals | Ensuring that net zero strategies are economically viable. | AI-driven optimization models that balance economic factors with environmental impact to identify cost-effective sustainability strategies. |

*4.3 Green washing model*

Greenwashing is a deceptive marketing strategy in which an FI exaggerates or fabricates the environmental benefits of its products, services, or practices to appear more environmentally friendly than they actually are. It means greenwashing pertains to the practice of misleadingly portraying a company as environmentally responsible, when in reality, it misleads investors and consumers. Some common behaviors include:

- Misleading Labels and Claims: Using vague, unregulated, or misleading terms like 'eco-friendly', 'green', or 'natural' without substantial evidence.
- Overstating Environmental Initiatives: Exaggerating the positive environmental impact of products or practices.
- Irrelevant Claims: Highlighting an environmentally friendly aspect which is a legal requirement.
- Hidden Trade-offs: Focusing on one environmentally friendly aspect while ignoring other significant environmental harms.

- Lack of Transparency: Withholding information or being vague about corporate environmental practices and impacts.

Assertions of offering "green products" must be substantiated with factual data and statistics. This ambiguity in distinguishing truly green initiatives from non-green ones partially drives the development of the EU Taxonomy. This regulation is designed to assist stakeholders in more objectively assessing a company's environmental sustainability and to simplify the reporting process for companies. The issue is further compounded by the use of vague terminology in environmental claims, where undefined or overly broad terms make it challenging to quantify and assess these claims accurately. Additionally, the complexity of global and multifaceted supply chains adds a significant hurdle in tracing and evaluating a company's environmental impact. This complexity is exacerbated by the subjectivity in determining what constitutes 'green' practices, leading to inconsistent assessments. Furthermore, the lack of standardized definitions and criteria for environmental friendliness across different industries and regions contributes to this inconsistency, complicating the evaluation process. Lastly, the vast amount of data and environmental claims that need to be analyzed and verified can lead to information overload, hindering the effective discernment of accurate and relevant environmental information from greenwashing. These factors collectively highlight the necessity for more rigorous, standardized, and transparent criteria in the evaluation of environmental claims to combat greenwashing effectively.

Table 12 outlines how AI can address each specific challenge associated with greenwashing, from ensuring the accuracy of environmental claims to managing the complexity and volume of relevant data. By integrating these AI-driven approaches, it becomes possible to tackle greenwashing more effectively and transparently.

*Table 12 AI solution for greenwashing*

| Challenge | Potential AI Solution |
| --- | --- |
| Substantiation of Green Claims | AI-driven data analytics to cross-reference and validate green product claims with empirical data. ML for analyzing large datasets. |
| Vague Terminology | NLP algorithms to interpret and analyze language in environmental claims, identifying vague or misleading terms. |
| Complexity in Supply Chains | AI algorithms for supply chain analysis, tracking, and evaluating environmental impact at each supply chain stage. |
| Subjectivity in Environmental Assessments | Development of AI models applying consistent criteria for 'green' practice assessments, reducing subjectivity. |
| Inconsistent Standards | AI systems adapting to and interpreting various environmental standards across industries and regions, providing unified assessment approaches. |
| Information Overload | Advanced AI algorithms for efficient data management and analysis, extracting and synthesizing relevant information from large data volumes. |

## 5. Responsible AI for ESG

Responsible AI is a cornerstone in the realm of ESG, playing a pivotal role in ensuring that technological advancements align with sustainable and ethical principles. The importance of responsible AI in ESG lies in its ability to enhance decision-making, optimize resource use, and drive innovation in a manner that is both equitable and environmentally conscious. Key considerations for implementing responsible AI in ESG include ensuring transparency in AI algorithms, which allows stakeholders to understand how AI makes decisions, particularly in critical areas like sustainable investment and risk assessment. Another vital aspect is the mitigation of biases, crucial for upholding fairness and preventing discrimination in AI-driven processes, such as hiring or lending practices. [32]

Ethical AI development is a key focus, requiring adherence to ethical standards and values from the inception to the deployment stages of AI systems. This includes respecting user privacy, ensuring data security, and aligning AI outcomes with broader societal values and ESG goals, such as

- Mitigating Biases and Ensuring Inclusivity: Actively work to identify and mitigate biases in AI algorithms to ensure fair and unbiased outcomes. This includes diversifying data sets and involving diverse teams in AI development to reduce the risk of unintentional discrimination.
- Promoting Transparency and Accountability: AI systems should be transparent in their operations and decision-making processes. Establish mechanisms for accountability to ensure that developers and users of AI are responsible for the outcomes of their systems. ESG transparency also eliminates communication barriers and improves the trustworthiness of the business among all its stakeholders.

More considerations and actions can be found in Table 13.

*Table 13 Responsible AI considerations and actions*

| Category | Consideration/Action | Detail/Example |
|---|---|---|
| Data/Model Privacy, Reliablity and Security | Implement model quality and robustness check | Use advanced cryptographic techniques to protect data at rest and in transit. |
| | Ensure compliance with data protection regulations | Adhere to GDPR of Europe, CCPA of US, PDPA of Singapore and other privacy laws for data handling and processing for the models. |
| | Use robust encryption | Use advanced cryptographic techniques to protect data at rest and in transit. |
| | Conduct regular security audits and monitoring | Perform vulnerability assessments and penetration testing to identify security gaps. Online (continuous) monitoring is an added-value. |
| | Adopt a privacy-by-design approach | Integrate data privacy into the design of AI systems, ensuring minimal data exposure. |
| | Enhance authorization and authentication | Utilize access control mechanisms to prevent unauthorized data breaches. |
| Accountability and Governance | Establish clear AI ethics guidelines | Create a code of conduct for AI development and usage reflecting ESG values. |
| | Assign responsibility for AI outcomes | Designate AI ethics officers to oversee responsible implementation and maintenance. |
| | Document AI decision-making processes | Maintain records of the logic, algorithms, and data used by AI systems for review and accountability. |
| | Develop transparent AI policies | Publicly share the organization's principles and standards for AI use in ESG contexts. |
| | Create a system for AI-related grievances | Set up channels through which stakeholders can report concerns or adverse AI impacts. |
| Model/Data Fairness and | Address and mitigate data biases | Use diverse datasets and algorithmic fairness techniques to prevent discriminatory AI outcomes. |

| Human-Centerity | | |
|---|---|---|
| | Perform impact assessments | Evaluate the potential consequences of AI deployment on various stakeholders. Determine the accepable thresholds and deviations |
| | Engage with relevant stakeholders and impacted communities | Include feedback from those affected by AI systems to guide equitable AI development. |
| | Prioritize accessibility and inclusivity | Ensure AI tools and platforms are accessible to a wide range of users with different abilities. |
| | Monitor for unintended consequences | Define clear metrics to measure the fairness and biases Continuously review AI systems post-deployment to identify and rectify unforeseen issues. |
| Transparency and Explainability | Develop explainable/interpreable AI models with clear data lineage and traceability | Use techniques and tools that allow for the interpretation of AI decision-making processes. The data shall be clearly traceable. |
| | Provide clear and understandable AI documentation and reporting mechanism | Make AI system documentation available and comprehensible for non-technical stakeholders. |
| | Facilitate third-party AI audits | Allow external evaluations of AI systems to verify their adherence to responsible AI practices. |
| | Implement a (human) feedback loop | Establish a mechanism for stakeholders to contribute to the continuous improvement of AI systems. |
| | Offer AI training and educational resources | Educate employees and stakeholders about how AI works and its role in the organization's ESG efforts. |

## 6. Sustainable AI

Sustainable AI can be viewed as an encompassing extension of the responsible AI paradigm, which traditionally focuses on the social implications of AI. This broader concept of Sustainable AI takes into account not only the societal impacts but also the environmental footprints of AI technologies with governance requirements. When FIs devise their Sustainable AI strategies, they must navigate a dual-focused pathway: one that leverages AI to advance sustainability goals, and another that scrutinizes the sustainability of AI systems themselves [1].

The facet of "AI for sustainability" involves deploying AI technologies as tools to further the ambitions of global sustainability efforts, such as the United Nations SDGs. Examples include using AI to optimize energy distribution networks, thereby reducing waste and promoting the use of renewable resources, or employing ML algorithms to analyze vast datasets for trends that can inform climate action strategies, as shown in Table 4 Environmental applications—initiatives often encapsulated under banners like AI4Good or AI4climate. These applications of AI have the potential to drive significant progress in addressing some of the world's most pressing environmental challenges.

On the flip side, "the sustainability of AI" calls for a critical examination of the energy consumption and resource use throughout the lifecycle of AI technologies—from the data centers that power computation-intensive tasks to the end-of-life disposal of hardware. FIs must adopt practices that increase the energy efficiency of AI operations, such as

o   utilizing more energy-efficient algorithms, e.g., use LLM only when necessary if the traditional ML algorithms cannot work well

o   leveraging and tuning existing models instead of training new models from scratch

o   Optimizing resource usage through DevOps automation, system auto scale, scheduling algorithm, and selection of cloud providers and data centers.

o   investing in green data centers powered by renewable energy sources.

Moreover, the governance of AI applications demands stringent oversight to protect various facets of ethical AI usage as discussed in the previous sectors, including but not limited to ensuring privacy through secure data practices, maintaining transparency in AI decision-making, enforcing accountability for AI-driven outcomes, upholding fairness to prevent algorithmic bias, and safeguarding justice by allowing recourse in the event of AI-induced harm.

In essence, Sustainable AI strategy within FIs is a comprehensive framework that aims not only to exploit AI's capabilities for promoting ecological and societal wellbeing but also to do so in a way that is itself sustainable and ethically sound [1] [33] [34] [35] [36].

## 7.   ESG Into Banking Process, in particular, Investing

Transforming a FI to embrace an ESG, e.g., Net Zero, strategy is a multifaceted endeavor that requires significant investment, mirroring the scale of other strategic corporate priorities. This transformation necessitates the integration of ESG considerations into every aspect of corporate finance functions, including operations, financial planning, business strategy, product development and pricing, credit assessment, facilities management, corporate venture investing, and treasury operations.

Use Net Zero as an example. To develop a coherent net zero corporate investment strategy, banks must begin by assessing their assets under ownership or control. This involves evaluating the environmental footprint of their loan portfolios, investments, and direct asset holdings. Concurrently, there is a need to analyze the value chain for potential opportunities and risks associated with net zero, ensuring that the bank's operations and financial offerings contribute positively towards environmental sustainability [31].

Creating a net zero investment portfolio requires collaboration with business strategy and (product) development teams, as well as sales departments. This collaboration aims to realize the potential of the net zero portfolio and understand its impacts on the market and the bank's financial health. Moreover, embedding net zero criteria into economic assessments and integrating these considerations across various financial functions, including Treasury, Risk, Audit, Mergers & Acquisitions, and strategic and financial planning, is crucial.

Another key aspect is identifying and prioritizing financing mechanisms and sources. This process entails evaluating both internal revenue generation and external financing options to support the bank's net zero-related transformation efforts. This could include innovative financial products like green bonds and sustainable loans, which are becoming increasingly important in the modern financial landscape.

Finally, the bank's approach to corporate venture capital and accelerators/incubators should also evolve to reflect its net zero commitment. This involves setting net zero as a strategic investment focus area, supporting startups and technologies that align with and advance the bank's sustainability goals. By doing so, the bank not only contributes to the broader environmental objectives but also positions itself as a leader in the financial sector's transition towards sustainability.

Therefore, the journey to net zero for a FI is requiring a strategic approach to asset management, value chain analysis, investment portfolio development, and the integration of sustainability into all facets of corporate finance. By addressing these key areas, banks can effectively navigate the transition to a net zero future, balancing their economic goals with environmental responsibility, and setting a precedent for sustainable practices in the financial industry. This transition not only aligns with global environmental targets but also opens up new avenues for growth and innovation in green finance, reinforcing the bank's commitment to a sustainable future.

Incorporating AI into sustainable investment strategies could significantly enhance investors' capacity to manage the intricate array of ESG factors. Leveraging the analytical power of AI, investors are empowered to pinpoint companies exhibiting robust ESG practices, reduce risks, and construct portfolios more closely aligned with sustainability goals. [37]

To effectively embed ESG requirements into core banking processes, particularly for investing, FIs must undertake a comprehensive strategy as mentioned earlier. This starts with integrating new AI-driven workflows into existing processes, enhancing decision-making with rich ESG data insights. The application of AI allows for more nuanced and data-driven decisions, aligning investments with ESG criteria. Collaboration with external ESG experts or consultants is required sometimes to gain insights and stay updated on best practices and emerging trends in sustainable investing.

Parallelly, an organization-wide communication of ESG mandates is essential, achieved through a strategic change management approach that brings all employees on board. This involves not just disseminating information but also engaging all employees through a well-thought-out change management strategy, enhanced by AI-driven communication tools. These tools can tailor and personalize messages for different employee groups, ensuring clearer understanding and alignment with ESG directives. Such an approach ensures that every member of the organization understands and aligns with the new ESG directives. The organization may create specialized ESG roles or teams dedicated to monitoring and guiding ESG integration within investment processes.

Regular reviews and revisions of data processes are critical to comply with changing ESG standards. Here, AI can be employed to automate the collection and processing of ESG data, increasing the frequency of data updates and enhancing accuracy and relevancy [1]. This might involve leveraging AI algorithms for predictive analytics to anticipate future ESG trends and standards, which are crucial for informed ESG decision-making. Robust ESG reporting mechanisms, powered by AI, shall be implemented to track progress and maintain transparency with stakeholders. These AI-driven reports can provide deeper insights and predictive analyses, enhancing stakeholder engagement. Regular audits of ESG integration efforts shall be conducted with the aid of AI tools to identify areas for improvement and ensure adherence to set standards through advanced pattern recognition and anomaly detection.

Additionally, developing a clear plan for integrating new ESG policies is crucial for seamless adoption. AI-powered project management and workflow automation tools can facilitate this integration, providing practical guidelines and tracking the implementation of new measures. This includes the addition of new certificates to investments, ensuring that these integrations are seamless and effective. ESG training programs, augmented by AI-driven learning platforms, shall be established for employees, particularly those involved in investment decisions. These platforms can offer personalized learning experiences and adaptive content to build a strong foundation in ESG principles and practices while using AI to track learning progress and adapt training materials to individual learning styles. Table 14 summarizes some essential changes in the processes, and we can see that AI can be applied almost all tasks as shown in Table 5 and Table 6.

*Table 14 ESG Changes in banking process*

| Banking Process | Change to be Implemented | Reason for Change | Potential AI Solution |
|---|---|---|---|
| Client Onboarding | Integrate ESG preference questionnaire. | To understand the client's inclination towards ESG-focused investments. | AI-driven survey analysis tools to interpret client responses and categorize clients based on ESG preferences. |
| Risk Assessment and Management | Include ESG risks in comprehensive risk assessment. | To identify and mitigate long-term risks associated with ESG issues. | AI algorithms for comprehensive risk analysis incorporating ESG factors, predictive modeling for long-term ESG risks. |
| Investment Analysis and Decision Making | Implement ESG scoring and rating systems. | To select investments that are financially sound and responsible. | Use AI to develop and manage dynamic ESG scoring systems, integrating them into existing investment analysis tools. |
| Portfolio Management | Prioritize assets with high ESG ratings. | To align portfolios with sustainable and ethical practices. | AI-based portfolio optimization tools that prioritize ESG-rated assets, aligning portfolios with sustainable practices. |
| Client Advisory Services | Offer advice on ESG investment options and impact. | To promote awareness and adoption of ESG investing among clients. | AI-powered recommendation engines to provide personalized ESG investment options based on client profiles and preferences. |
| Performance Monitoring and Reporting | Incorporate ESG performance metrics in reporting. | To evaluate the social and environmental impact alongside financial returns. | AI for data aggregation and analysis, generating reports with integrated ESG performance metrics alongside financial returns. |

## 7. Concluding Remarks with Collaborative Paths

The integration of AI into sustainability and ESG practices presents a dual landscape of opportunities and challenges. While AI holds immense potential to drive positive change through resource optimization, renewable energy management, climate action, and transparent ESG reporting, it also demands vigilant efforts to address data privacy, bias, ethical concerns, and accessibility barriers. By fostering collaboration, ethical development, stakeholder engagement, and sustained research, society can chart a path forward where AI serves as a powerful tool in building a more sustainable and equitable future.

To effectively harness AI's transformative power in the context of ESG, while minimizing its potential drawbacks, a collaborative approach is required using the Shared Value Model [38]. In the Shared Value Model, companies are recognized as a key part of the solution to sustainable development rather than being seen as a problem. This perspective holds that sustainable practices can, over time, enhance a corporation's profitability. Yet, within this framework, it is vital to concentrate on the appropriate matters related to the company's impact. This approach should encompass the following strategic elements, each accompanied by specific actions and examples:

- Building Global Cooperation and Industrial Standardization: Establish global cooperation to develop international AI standards and best practices. This approach addresses cross-border AI challenges and ensures uniform governance and ethical considerations, akin to the international collaboration seen in climate change agreements.

- Building Resilient and Adaptive Regulatory Frameworks: Develop regulatory frameworks by regulators and industrial associations that can keep pace with AI advancements, ensuring responsible and ethical usage. For instance, creating policies that adapt to new AI technologies in financial markets, ensuring they remain fair and transparent.

- Establishing Ethical Guidelines and Standards: Formulate and follow ethical guidelines for AI, incorporating transparency, fairness, privacy, and accountability. These guidelines would ensure AI in healthcare respects patient confidentiality and delivers unbiased treatment recommendations.

- Fostering Public-Private Partnerships: Encourage collaboration between governments, private sectors, and academia for a balanced AI development approach. Such partnerships might result in joint ventures for developing AI in sustainable energy, combining academic research with industrial application.

- Enhancing Data Governance: Implement strong data governance policies to maintain data integrity and security, like setting strict data handling protocols for AI applications in banking to protect customer data.

- Investing in AI Literacy and Education: Due to the intricate nature of sustainable development, the adoption and integration of new economic models typically involve incorporating various disciplines, which require solid literacy and education. Enhance AI understanding among all stakeholders through education and training programs. For example, offering AI ethics courses in universities and corporate training programs.

- Encouraging Stakeholder Engagement: Involve a broad spectrum of stakeholders, including underrepresented communities, in AI development and deployment. This could include community-led AI projects in urban development to ensure technologies meet the real needs of residents.

- Conducting Deep Research and Development: Continuously invest in research to understand AI's societal, environmental, and industrial impacts. This involves initiatives like funding long-term studies on AI's effects on job markets and developing AI solutions for environmental conservation.

Each of these elements plays a crucial role in ensuring that AI is developed and utilized in a collaborative way that maximizes its benefits for ESG initiatives, while carefully managing and mitigating any potential risks or negative impacts.

Another key point worthy of mentioning in the context of utilizing AI for ESG and sustainability, it's equally important to focus on developing sustainable AI technology. This approach aims to minimize any adverse effects on ESG factors, as well as on IT infrastructure. [34] [35] [33] [36]

*Figure 3 Reference architecture of AI-powered ESG platform*

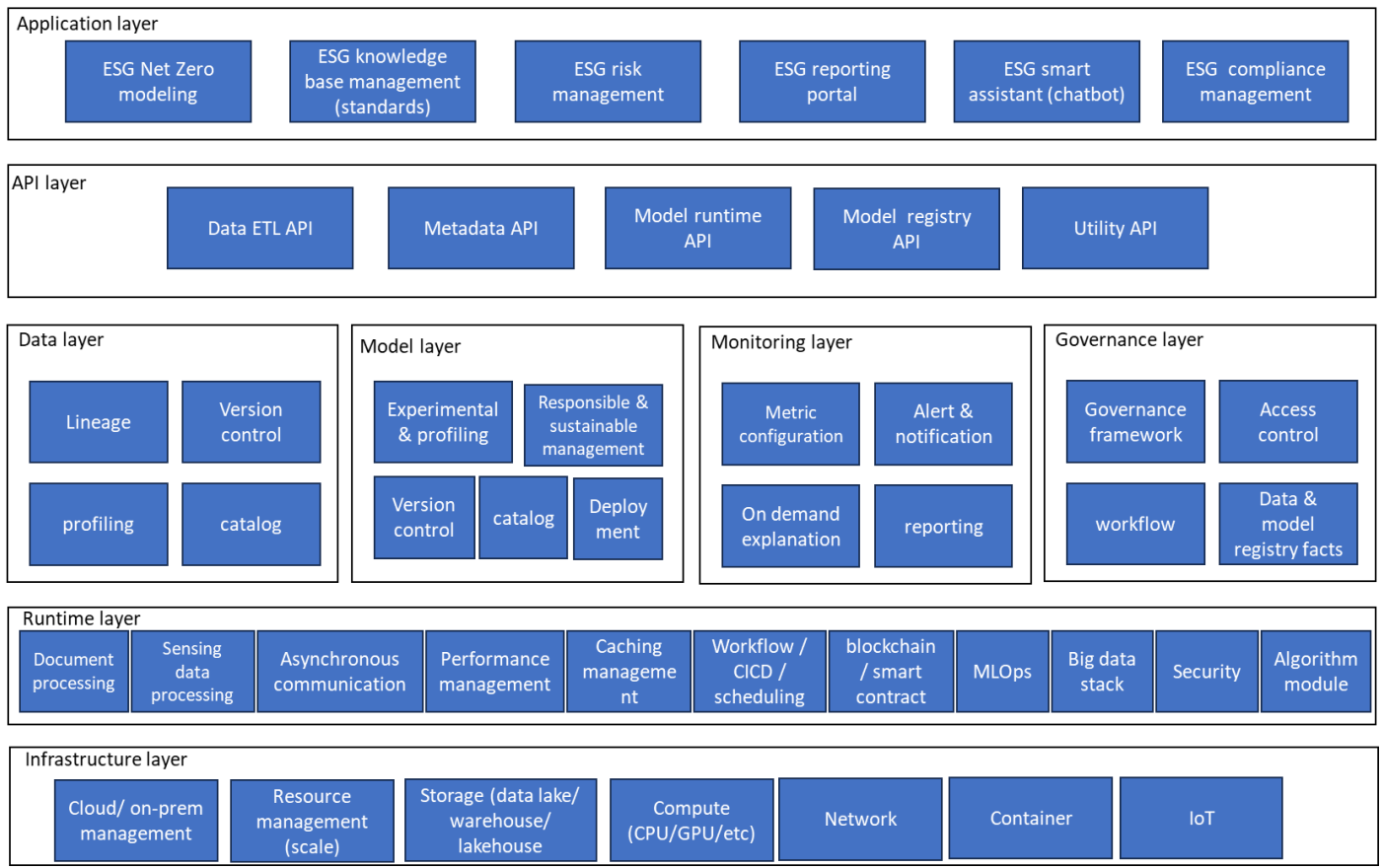

In conclusion, it is essential to recognize that the sophistication of AI applications in the financial sector is contingent upon their harmonious integration with other financial technologies (FinTech), including blockchain, cloud computing, big data, and the Internet of Things (IoT). Reference [1] elucidates this synergy. Figure 3 delineates a reference architecture for an ESG platform, where the foundational infrastructure layer is bolstered by the robust capabilities of various FinTech solutions, and run-time layer is built by AI tool, MLOps automation and general management components.

A cross-sectional examination of an enterprise-grade ESG platform reveals the implementation of standardization protocols at multiple tiers, encompassing data, modeling, monitoring, governance, and application programming interfaces (APIs). The data and model layers benefit from rigorous standardization through version management, cataloging, profiling, and classification, ensuring consistency and reliability. Similarly, the monitoring layer, which includes metrics monitoring, alerts, and notifications, as well as the provision of on-demand explanations, mandates uniformity across disparate systems for effective operation. The governance layer helps to control the data and models' authentication and authorization, as well as the governance workflow to approve the usage of data and models. Within the API layers, the need for compliance becomes apparent in elements such as Data Connection APIs, ETL (Extract, Transform, Load) APIs, and Model Facts APIs, which are instrumental in upholding proper governance standards. Governance workflows further reinforce standardization by regulating access control and model facts documentation.

At the apex, the application layer forms the bridge connecting the "theoretical" frameworks of ESG platforms with practical ESG scenarios in the real world, demonstrating the tangible impacts of these technologies. This integration ensures that AI not only advances financial services but also aligns with the broader objectives of ESG criteria.

**Acknowledgments:** The author would like to thank Eva Zhao for the suggestion on Earthshot projects.

**Conflicts of Interest:** This paper only represents the options from the authors and doesn't represent any other organizations.

# References

[1]   J. Xu and et al, The Future and FinTech: ABCDI and Beyond, J. Xu, Ed., World Scientific Press, 2022.

[2]   Bloomberg, "ESG May Surpass $41 Trillion Assets in 2022, But Not Without Challenges, Finds Bloomberg Intelligence," Bloomberg, 24 January 2022. [Online]. Available: https://www.bloomberg.com/company/press/esg-may-surpass-41-trillion-assets-in-2022-but-not-without-challenges-finds-bloomberg-intelligence/.  [Accessed 19 December 2023].

[3]   PWC, "SG-focused institutional investment seen soaring 84% to US$33.9 trillion in 2026, making up 21.5% of assets under management: PwC report," PWC, 10 Octorber 2022. [Online]. Available: https://www.pwc.com/gx/en/news-room/press-releases/2022/awm-revolution-2022-report.html. [Accessed 19 December 2023].

[4]   "Retail investor capital presents USD8.2 trillion growth opportunity for sustainable investing," Standard Chartered Bank, 26 September 2022. [Online]. Available: https://www.sc.com/en/press-release/retail-investor-capital-presents-usd-8-2-trillion-growth-opportunity-for-sustainable-investing/. [Accessed 24 December 2023].

[5]   O. Alexander and D. Yazdani, "Exponential Expectations for ESG," Harvard Law School Forum on Corporate Governance, 17 November 2022. [Online]. Available: https://corpgov.law.harvard.edu/2022/11/17/exponential-expectations-for-esg/. [Accessed 19 December 2023].

[6]   "Morgan Stanley Survey Finds Investor Enthusiasm for Sustainable Investing at an All-Time High," Morgan Stanley, 12 September 2019. [Online]. Available: https://www.morganstanley.com/press-releases/morgan-stanley-survey-finds-investor-enthusiasm-for-sustainable-. [Accessed 24 December 2023].

[7]   C. D. Ditlev-Simonsen, A Guide to Sustainable Corporate Responsibility: From Theory to Action, Switzerland: Palgrave Macmillan, 2022.

[8]   "Poseidon Principles," [Online]. Available: https://www.poseidonprinciples.org/finance/. [Accessed 19 Janurary 2024].

[9]   E. White and et al., "THE SBTi FINANCIAL INSTITUTIONS NETZERO STANDARD: CONCEPTUAL FRAMEWORK AND INITIAL CRITERIA," SBTi, 2023.

[10]  "Facilitated Emissions: Global GHG Accounting and Reporting Standard (Part B)," PCAF, 2023.

[11]  A. Saxena and et al., "Technologies Empowered Environmental, Social, and Governance (ESG): An Industry 4.0 Landscape," *sustainability,* vol. 15, no. 309, p. 17, 2023.

[12]  L. Cao, "AI in Finance: Challenges, Techniques and Opportunities," *ACM Computing Surveys,* vol. 12, no. 2, pp. 1-38, 2022.

[13]  L. Cao, "AI in Finance: A Review," *SSRN Electronic Journal,* p. 36, 2020.

[14]  G. Pisoni, B. Molnár and Á. Tarcsi, "Data Science for Finance: Best-Suited Methods and Enterprise Architectures," *Applied System Innovation,* vol. 4, no. 69, p. 20, 2021.

[15]  E. Burnaev, E. Mironov, A. Shpilman, M. Mironenko and D. Katalevsky, "Practical AI Cases for Solving ESG Challenges," *sustainability,* vol. 15, no. 12731, p. 15, 2023.

[16]  "ML system design: 300 case studies to learn from," Evidently AI, 2 December 2023. [Online]. Available: https://www.evidentlyai.com/ml-system-design?utm_source=talkingdev.uwl.me. [Accessed 19 December 2023].

[17] Microsoft News Center, "Microsoft, Planet and The Nature Conservancy launch the Global Renewables Watch," Microsoft, 22 September 2022. [Online]. Available: https://news.microsoft.com/2022/09/22/microsoft-planet-and-the-nature-conservancy-launch-the-global-renewables-watch/. [Accessed 19 December 2023].

[18] Y. Pilankar, R. Haque, M. Hasanuzzaman, P. Stynes and P. Pathak, "Detecting Violation of Human Rights via Social Media," in *Proceedings of CSR-NLP I @LREC*, Marseille, 2022.

[19] "The Earthshot Prize," 2023. [Online]. Available: https://earthshotprize.org/winners-finalists-listing. [Accessed 13 Janurary 2024].

[20] R. Ghous, "How a Modern Data Architecture Brings AI to Life: Data Mastering for AI," Informatica, 19 Octorber 2023. [Online]. Available: https://www.informatica.com/blogs/how-a-modern-data-architecture-brings-ai-to-life-data-mastering-for-ai.html. [Accessed 19 December 2023].

[21] D. Heller, S. Schöbl, H. Soller and A. Reiter, "ESG data governance: A growing imperative for banks," McKinsey, 3023.

[22] TDFD. [Online]. Available: https://tdfd-global.org/.

[23] A. Spezzatti, E. Kheradmand, K. Gupta, M. Peras and R. Zaminpeyma, "Note: Leveraging Artificial Intelligence to build a Data Catalog and support research on the Sustainable Development Goals," in *COMPASS '22: Proceedings of the 5th ACM SIGCAS/SIGCHI Conference on Computing and Sustainable Societies*, 2022.

[24] L. Kelly, "What is ESG Data? Uses, Types & Dataset Examples," Datarade, 13 December 2023. [Online]. Available: https://datarade.ai/data-categories/esg-data. [Accessed 19 December 2023].

[25] Y. Li, S. Wang, H. Ding and H. Chen, "Large Language Models in Finance: A Survey," in *4th ACM International Conference on AI in Finance (ICAIF-23)* , 2023.

[26] H. Zhao and et al., "Revolutionizing Finance with LLMs: An Overview of Applications and Insights," arxiv, Janurary 2024. [Online]. Available: https://arxiv.org/pdf/2401.11641.pdf. [Accessed 28 Janurary 2024].

[27] "RepRisk methodology overview," RepRisk, 2021.

[28] "ESG Data and Analytics from Truvalue Labs," TRUVALUE LABS, 2021.

[29] "Market-Leading ESG Solutions," ESGbook, 2021. [Online]. Available: https://www.esgbook.com/analytics-solutions/. [Accessed 19 December 2023].

[30] "Standard Chartered NetZero Approach: Methodological white paper," Standard Chartered, 2022.

[31] M. Azoulay , A. Casoli, D. Mikkelsen , M. Muvezwa, D. Stephens, S. Venugopal, S. Underwood and D. Yang, "Managing financed emissions: How banks can support the net-zero transition," McKinsey & Company, 2022.

[32] S. Mazumder, S. Dhar and A. Asthana, "A Framework for Trustworthy AI in Credit Risk Management: Perspectives and Practices," *Computer* , vol. 56, no. 5, pp. 28 - 40, 2023.

[33] A. van Wynsberghe, "Sustainable AI: AI for sustainability and the sustainability of AI," *AI and Ethics,* vol. 1, p. 213–218, 2021.

[34] V. Galaz and et al, "Artificial intelligence, systemic risks, and sustainability," *Technology in Society,* vol. 67, no. 101741, 2021.

[35] C. Isensee, K.-M. Griese and F. Teuteberg, "Sustainable artificial intelligence: A corporate culture perspective," *NachhaltigkeitsManagementForum,* vol. 29, p. 217–230, 2021.

[36] I. S. Banipal, S. Asthana and S. Mazumder , "Sustainable AI - Standards, Current Practices and Recommendations," in *Proceedings of the Future Technologies Conference*, 2023.

[37] "How AI Can Bolster Sustainable Investing," morgan stanley, 31 July 2023. [Online]. Available: https://www.morganstanley.com/ideas/ai-sustainable-investing-use-potential. [Accessed 19 December 2023].

[38] M. E. Porter and M. R. Kramer, "Creating shared value: How to reinvent capitalism—and unleash a wave of innovation and growth," *Harvard Business Review,* vol. 89, no. 1-2, p. 62, 2011.